\documentclass[useAMS,usenatbib]{mn2e}

\usepackage {graphicx}

\title[Barium Abundances in Cepheids]
{Barium Abundances in Cepheids}
\author[S.M.~Andrievsky et al.]
{S.M.~Andrievsky$^{1,2,3}$\thanks{E-mail:scan@deneb1.odessa.ua},
J.R.D.~L\'epine$^{1}$, 
S.A.~Korotin$^{2}$, 
R.E.~Luck$^{4}$,
  \newauthor 
V.V.~Kovtyukh$^{2}$,
and W.J.~Maciel$^{1}$
\\
$^{1}$Instituto de Astronomia, Geof\'{\i}sica e Ci\^encias Atmosf\'ericas da
   Universidade de S\~ao Paulo,\\ Cidade Universit\'aria, CEP: 05508-900, 
   S\~ao Paulo, SP, Brazil\\
$^{2}$
   Department of Astronomy and Astronomical Observatory, Odessa
   National University,\\ 
   and Isaac Newton Institute 
   of Chile Odessa branch Shevchenko Park, 65014 Odessa, Ukraine.\\
$^{3}$
   GEPI, Observatoire de Paris-Meudon, F-92125 Meudon Cedex, France,\\
$^{4}$
  Department of Astronomy, Case Western Reserve University, Cleveland, OH 44106.\\
}

\begin{document}

\date{Accepted. Received ; in original form }

\pagerange{\pageref{firstpage}--\pageref{lastpage}} \pubyear{2011}

\maketitle

\label{firstpage}

\begin{abstract}

We derived the barium atmospheric abundances for a large sample
of Cepheids, comprising 270 stars. The sample covers a large
range of galactocentric distances, from about 4 to 15 kpc,
so that it is appropriated to investigate the existence of 
radial barium abundance gradients in the galactic disc. In fact, 
this is the first time that such a comprehensive analysis of the 
distribution of barium abundances in the galactic disc is carried
out. As a result, we conclude that the Ba abundance distribution
can be characterized by a zero gradient. This result is compared
with derived gradients for other elements, and some reasons
are briefly discussed for the independence of the barium abundances
upon galactocentric distances.

\end{abstract}


\section{Introduction}  

In our extensive studies of the elemental distributions in the galactic disc based on studies of 
Cepheid spectra (\citealt{An02a}, \citealt{An02b}, \citealt{An02c}, \citealt{An04}, \citealt{Lu03}, 
\citealt{Lu06}, \citealt{Lu11}, \citealt{LL11}), we have avoided the determination of barium abundances.  
Although three Ba\,{\sc ii} lines are available in the observed spectral domain, all are very strong 
in Cepheid spectra with equivalent widths ranging from 200 to 600~m\AA. A simple LTE analysis of 
such strong lines could produce incorrect abundance results due to NLTE effects in the atomic 
populations, and more importantly, due to strong saturation effects making the analysis 
particularly sensitive to the microturbulence. 

The great majority of studies devoted to the barium abundance in the galactic disc are based 
on the LTE approximation, and most focus on nearby stars. Recently, a more sophisticated 
method: i.e., NLTE, has been applied to the galactic disc stars, but as in the case of older 
studies, only for those situated near the solar region (\citealt{Ko10}).

The present-day barium abundance distribution in galactic disc results from a variety of 
processes.  The different processes include nucleosynthesis in AGB stars, supernovae of type 
II, as well as dynamical processes involving radial gas flows that lead to large scale mixing of 
the interstellar medium.  The barium abundance distribution in the galactic disc (as well as 
distributions of the heavy $r$-/$r+s$-process elements like La, Ce and Nd) is necessary to 
constrain models of the Milky Way chemical evolution. Existing data on the barium content in 
objects situated at large distances from the Sun are scarce, and apparently insufficient to 
check the validity of model predictions (see, e.g. Figure 9 in \citealt{Ce07}).

To investigate the barium abundances in distant stars we use the sample of Cepheids 
investigated in our previous works: \citet{An02a}, \citet{An02b}, \citet{An02c}, 
\citet{Lu03}, \citet{Lu06}, \citet{Lu11}. In those papers, one can find information about the 
atmospheric parameters of the program stars and their spectra.  For some stars, we have used 
the multi-phase observations described in \citet{Lu04}, \citet{Ko05}, \citet{An05}, and \citet{Lu08}.  
The reasons for studying these stars are many.  These stars are luminous, allowing them to serve as 
probes of distant regions. The methods of the abundance analysis of the stars of F-G spectral 
classes are well established; and finally, Cepheids are rather young stars, and thus they reflect 
the present-day characteristics of the galactic disc, even though they had time to migrate. 

\section{Barium atomic model and method of abundance determination}.

\citet{An09} describe our model of barium atom in detail.  Briefly, it consists of 31 
levels of Ba\,{\sc i}, 73 levels of Ba\,{\sc ii}, and the ground level of Ba\,{\sc iii}. 
Ninety-one bound-bound transitions between the first 28 levels of Ba\,{\sc ii} with $n<$ 12 
and $l<$ 5 are computed in detail. The remaining levels are used for particle number conservation. 
For two levels, 5d${}^2$D and 6p${}^2$P$_{0}$, fine structure was taken into account. Oscillator 
strengths, photoionization cross-sections, collisional rates, broadening parameters and test 
calculations are found in the above-mentioned paper. 

Atomic level populations were determined using the MULTI code of \citet{Ca86} with 
modifications as given in \citet{Ko99}. MULTI calculates the line profile for each line 
considered in detail. The line profile computed assuming either LTE or NLTE depends upon 
many parameters: the effective temperature of the model, the surface gravity, the 
microturbulent velocity, and the line damping as well as the populations in the appropriate 
levels.  Departure coefficients, defined as the ratio of NLTE to LTE level populations, are also 
computed by MULTI, and they depend only on the model atom and model stellar atmosphere.  
Specifically, they are independent of the total abundance, the damping, and the 
microturbulence.

Barium atom has seven isotopes. Isotopic shifts are very small (about 2 m\AA).
For two odd isotopes ${}^{135}$Ba and ${}^{137}$Ba the hyper-fine structure is quite 
important, and this produces additional complication in the line structure 
(several line components of the odd isotopes shifted relatively to the lines of even 
isotopes). This produces the shifts between the line components formed by odd and even
isotopes. It is known that the most pronounced effect is seen in the Ba\,{\sc ii} 
line 4554.03 \AA ~(out of our spectral region).

In principle, this effect has some influence on profile of the 6496.91 \AA~line.
The components of the odd isotopes constitute two compact groups that are
shifted on --4 and +9 m\AA~ relatively components of the even isotopes. For 
the rest of our two lines the corresponding shifts are too small to affect 
their line profiles.

As it was showed in the work of \citet{Mashet99} for the adequate barium 
line modelling it is sufficient to use the three-component model suggested by 
\citet{Rut78}. For the calculations of the Ba line profiles in the spectra 
of young stars one can use the even-to-odd abundance ratio of 82:18 
(\citealt{Cam82}).

Since the equivalent widths of our program barium lines are larger than 
250 m\AA, we can state that HFS does not have in this case any significan 
influence on the line profiles.

\begin{figure}
\resizebox{\hsize}{!}
{\includegraphics {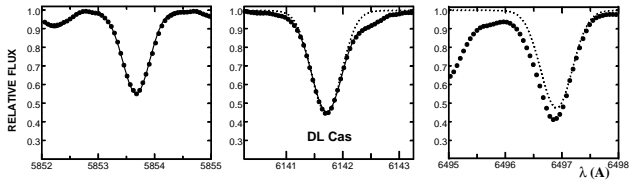}}
\caption[]{Profile fitting of the Ba\,{\sc ii} line in the DL~Cas spectrum. Observed  spectrum - dots,  
pure NLTE MULTI profile - dotted  line, combined synthetic spectrum - smooth line.}
\label {DL Cas}
\end{figure}

\begin{figure}
\resizebox{\hsize}{!}
{\includegraphics {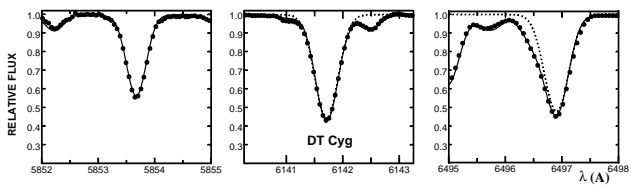}}
\caption[]{The same as Figure 1 but for DT Cyg.}
\label {DT Cyg}
\end{figure}

Three Ba\,{\sc ii} lines are available in our program spectra for the abundance analysis: 5853.68~\AA, 
6141.71~\AA~ and 6496.91~\AA. The barium abundance was derived by fitting calculated profiles to 
the observed profiles. All of the available barium lines are blended with lines of other elements. 
(iron lines, in particular). The blending effect is more significant for the 6141.71 and 6496.91~\AA~ 
lines (for 5853.68~\AA line this effect is negligible). Proper comparison of the observed and computed 
profiles thus requires a multi-element synthesis.  For this process, we fold the NLTE (MULTI) calculations 
into the LTE synthetic spectrum code SYNTHV (\citealt{Ts96}). With these programs, we calculated synthetic 
spectra for each Ba\,{\sc ii} line region taking into account all the lines in each region listed 
in the VALD database (http://ams.astro.univie.ac.at/vald/). For the barium lines, the corresponding 
departure coefficients (so-called $b$-factors: $b = n_{\rm i}/n^{*}_{\rm i}$ - the ratio of 
NLTE-to-LTE level populations) are input to SYNTHV, where they are used in the calculation of the 
line source function and barium line profiles. Figures \ref{DL Cas} and \ref{DT Cyg} show the quality 
of fit achieved by this process.  

\begin{figure}
\resizebox{\hsize}{!}
{\includegraphics {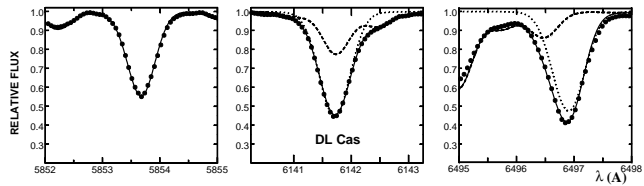}}
\caption[]{Ba~II lines, blending lines and resulting profiles in DL~Cas spectrum.
Dots - observed spectrum, smooth line - combined calculated profile, dotted line - 
pure Ba\,{\sc ii} line profile, dashed line - blending spectral line.}
\label {Blend}
\end{figure}

In Fig. \ref{Blend}  we show how blending lines together with Ba\,{\sc ii} lines form the resulting profile 
of 6141.71 and 6496.91 \AA~lines (an example for Cepheid DL~Cas). We can see that the contribution 
from the blending lines (Fe\,{\sc i} 6141.732 \AA~ for  6141.71 \AA, and  Fe\,{\sc i} 6496.467 \AA~ 
for 6496.91 \AA) is not extremely large (we have also to note that there are also two weak lines in 
the wings of 6141.71 \AA~ line - Fe\,{\sc ii} 6141.033 \AA~ and Si\,{\sc i} 6142.483 \AA). If spectral 
synthesis is properly done, then this contribution is adequalely accounted. For blending lines 
we used the reliable atomic data from the VALDatabase, see the reference above, while abundances of 
elements producing those lines are taken from the series of our previous papers, i.e. they were fixed. 
In particular, in this case VALD data are based on original data from \citet{FurWi06}.
In fact, we always relied more on the barium abundance derived from 5853.68 \AA~line. In a few cases 
this line was not available in our spectra, or its profile was spoiled due to some reasons. In those 
cases, only the rest two lines were used in abundance analysis. As a rule, the barium abundance derived 
from 5853.68 \AA~ line and from other two lines agreed well between themselves. If there was a significant 
deviation of the barium abundance from one of the 6141.71 and 6496.91 \AA~lines, the correponding 
line was not considered in the final statistics.

Stellar parameters for all stars were taken from our previous works (for a compendium, see 
\citealt{Lu11}). These parameters were determined from a combined LTE excitation and 
ionization balance analysis.  In particular, the microturbulent velocity depends solely on Fe\,{\sc ii}
lines. In our syntheses, we used previously published (see \citealt{Lu11}) abundances of elements 
whose lines were treated in SYNTHV: i.e., blending lines and lines situated in the vicinity of barium 
lines.

In some cases, our program Cepheids exhibit asymmetric line profiles due to dynamic 
phenomena associated with pulsation. In such cases, our profile fitting cannot be applied, and 
thus, spectra with greatly asymmetric lines were excluded from the analysis. A few exceptions 
were allowed in which we used the 5853.68~\AA~line alone to derive NLTE barium abundance. 
This line is not heavily blended, and its equivalent width is measurable by direct integration. 

\section{Results of the NLTE barium abundance determination and sources of abundance errors}.

Table \ref{tabstars} contains the list of program stars, phases for the selected spectra, adopted 
atmosphere parameters, galactocentric distances (based on R$_{\rm Sun} = 7.9$ kpc), NLTE 
barium abundances, sigma values and the number of analyzed barium line profiles. In some cases, 
adjacent echelle orders contain the same barium lines and each profile was analyzed 
independently. For some stars we have multiphase observations, therefore the number of 
analyzed barium line profiles is the number of available spectra multiplied by the number 
of available barium lines in each spectrum (this number is listed in Table \ref{tabstars}). 

We have investigated the parameter sensitivity of the derived Ba abundance.  Our initial parameters 
were T$_{\rm eff} = 6000$~K, $\log~g = 2.0$, V$_{\rm t} = 4.0$ km~s$^{-1}$ with [Fe/H] = 0.  
We then generated lines assuming $\log \epsilon$(Ba) = 2.17 and matched them using parameter 
sets that vary one parameter at a time.  An increase of 150~K in effective temperature changes 
the Ba abundance by +0.1 dex, a 0.2 dex increase in $\log~g$ gives a change of +0.06, and a 
microturbulent velocity increase and decrease of 0.5 km~s$^{-1}$ yields a decrease and increase 
in abundance of --0.23 and +0.28  dex respectively. The formal parameter related uncertainty 
is thus about $\pm$ 0.3 dex. However, note that at lower microturbulent velocities the dependence 
of the barium abundance on V$_{\rm t}$ is much stronger. 

\begin{figure}
\resizebox{\hsize}{!}
{\includegraphics {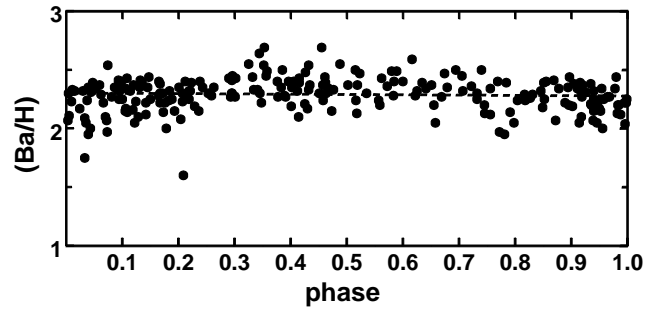}}
\caption[]{Barium abundance vs. phase for program stars spectra. Combined plot.}
\label {Ba_ph}
\end{figure}

\begin{figure}
\resizebox{\hsize}{!}
{\includegraphics {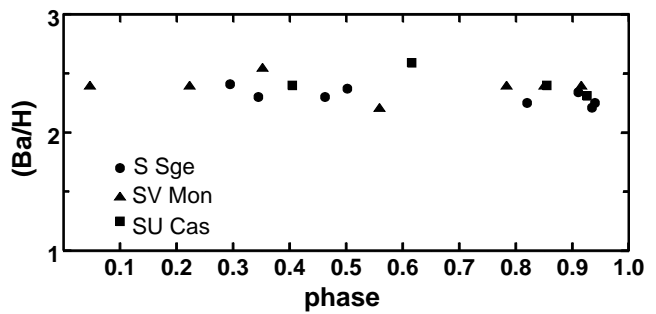}}
\caption[]{Barium abundance vs. phase for program stars spectra with multiphase
observations.}
\label {Ba_ph_ind}
\end{figure}

\begin{figure}
\resizebox{\hsize}{!}
{\includegraphics {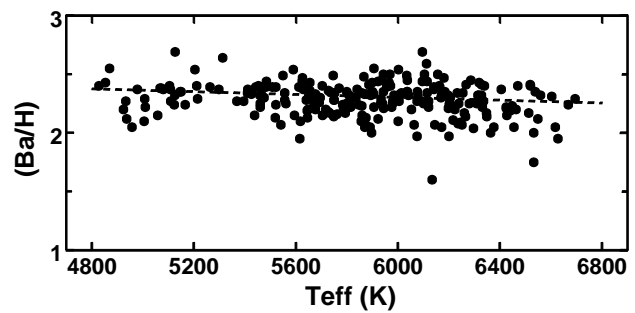}}
\caption[]{Barium abundance vs. effective temperature.}
\label {Ba_T}
\end{figure}

\begin{figure}
\resizebox{\hsize}{!}
{\includegraphics {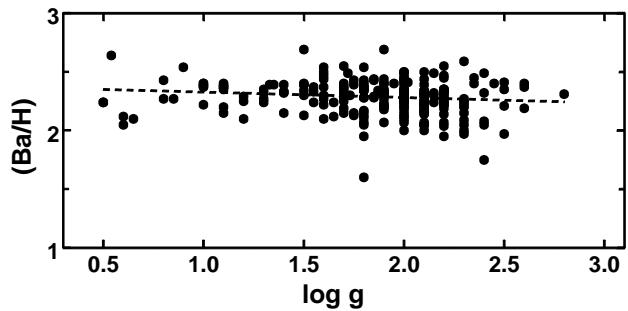}}
\caption[]{Same as Figure 6, but for surface gravity.}
\label {Ba_g}
\end{figure}

\begin{figure}
\resizebox{\hsize}{!}
{\includegraphics {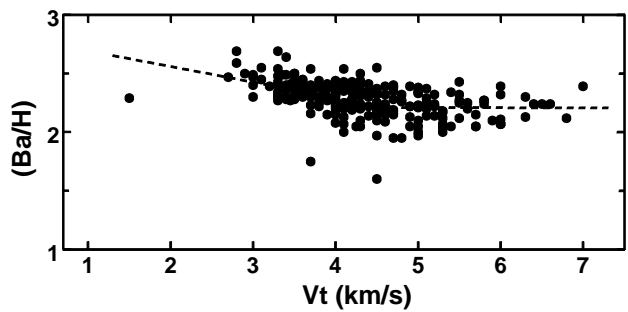}}
\caption[]{Same as Figure 6, but for microturbulent velocity.}
\label {Ba_V}
\end{figure}

In Figures \ref{Ba_ph} through \ref{Ba_V} we show the derived barium abundances as a function of the 
pulsation phase, effective temperature, surface gravity, and microturbulent velocity. As one can see, 
there is no significant dependence on phase $\phi$ (for three stars where we used multiphase observations 
Figure \ref{Ba_ph_ind} shows such a depenedence), T$_{\rm eff}$, or $\log~g$, but there is a quite 
clear variation with microturbulent velocity (Figure \ref{Ba_V}). This relation is quite disconcerting 
as it means that we are dealing with strong saturation in the lines, and hence, the dramatic dependence 
of abundance on assumed microturbulent velocity. To illustrate, we give in Table \ref{Tab2} the dependence 
of the Ba abundance on microturbulent velocity for UX~Car.

\begin{table}
\caption[]{NLTE abundances for UX Car}
\label{Tab2}
\begin{tabular}{ccccc}
\hline
\multicolumn{1}{c}{}&
\multicolumn{4}{c}{V$_{\rm t}$, km s$^{-1}$}\\
\hline
Line &   3.5 &  4.0 &   4.7 &   5.3 \\
\hline
5853 &   2.58&  2.37&   2.19&   2.09\\
6141 &   2.84&  2.53&   2.19&   1.99\\
6496 &   2.81&  2.50&   2.19&   2.01\\ 
\hline
\end{tabular}
\end{table}

In the case of UX~Car (Table \ref{Tab2}), the Fe\,{\sc ii} derived V$_{\rm t}$ (4.7 km~s$^{-1}$) yields excellent agreement 
between the three Ba\,{\sc ii} lines. This is not surprising as this V$_{\rm t}$ is in the region of 
V$_{\rm t}$ versus Ba (see Figure \ref{Ba_V}) where there is no obvious dependence between abundance and V$_{\rm t}$.  
This suggests that it might be possible to find the proper V$_{\rm t}$ value for Ba\,{\sc ii} by forcing the three 
lines to yield the same Ba abundance using V$_{\rm t}$ as a free parameter.  The result of this 
approach for IR~Cep is shown in Table \ref{Tab3}.  The idea behind this attempt is that each Ba\,{\sc ii} 
feature has a constant contribution from Ba that is defined by the best fitting synthesis.  
One then uses the best fitting abundance to the profile to calculate the unblended Ba equivalent 
width. One then matches the equivalent width for each selected microturbulent velocity.  
In this test, we have matched the equivalent widths using an LTE approach. As the primary 
problem we are trying to overcome now is not level populations but saturation effects.  While 
there is a range of equivalent widths in the Ba\,{\sc ii} lines, all lines are saturated and it is not 
possible to obtain a clear-cut indication of a better value to use for V$_{\rm t}$.  In this case, 
the Fe\,{\sc ii} V$_{\rm t}$ is 2.9 km~s$^{-1}$, which gives a spread of about 0.13 dex between the Ba\,{\sc ii} lines.  
The spread at 4.0 km~s$^{-1}$ is marginally better at 0.10 dex, but given the overall uncertainties, 
does not significantly improve the confidence level of the abundance determination. It is also of 
interest that above about 6 km~s$^{-1}$ the lines become only weakly dependent on V$_{\rm t}$; 
that is, they are de-saturated and behave as weak-lines.  Also included in Table \ref{Tab3} is the best fitting Ba 
abundance for the NLTE analysis.  The NLTE corrections for these lines are not especially large 
averaging around --0.1 dex.

\begin{table*}
\caption[]{Abundances for IR Cep}
\label{Tab3}
\begin{tabular}{ccccccccccc}
\hline
\multicolumn{2}{c}{}&
\multicolumn{9}{c}{V$_{\rm t}$, km s$^{-1}$}\\
\hline
Line & EW, m\AA &   2.0 & 2.5 & 3.0& 3.5 &  4.0 &4.5  &  5.0 &6.0 & 7.0 \\
\hline
LTE\\
\hline
5853 &196      &   3.23 & 2.81&  2.46&  2.18&  1.97&  1.83 & 1.73 &  1.58&  1.50\\
6141 &270      &   3.17 & 2.89&  2.54&  2.22&  1.94&  1.72 & 1.53 &  1.26&  1.09\\
6496 &280      &   3.30 & 2.96&  2.59&  2.29&  2.04&  1.83 & 1.67 &  1.43&  1.28\\
\hline
NLTE   &  &  &     2.75 &  &  2.08  \\
\hline
\end{tabular}
\end{table*}

The origin of the dependence of the Ba abundance of V$_{\rm t}$ as shown in Figure \ref{Ba_V} could arise 
in a velocity stratification in the atmosphere of Cepheids. The solar atmosphere exhibits an 
increasing microturbulent velocity with height (\citealt{St06}) and it would not be 
surprising if similar temperature but higher luminosity objects showed the same phenomena.    
Quite strong barium lines have effective depth of formation high in the atmosphere, while 
weaker (on average) iron lines used for microturbulence determination must be formed 
deeper in atmosphere at a lower V$_{\rm t}$.  

Since there is no straightforward method to account for the dependence seen in Figure \ref{Ba_V}, we 
simply divide the barium abundance data into two parts based on the microturbulent velocity. From 
Figure \ref{Ba_V}, one can estimate that the break in the (Ba/H)-V$_{\rm t}$ relation roughly corresponds 
to V$_{\rm t}$ = 3.8 km~s$^{-1}$. 

\section{The barium radial abundance distribution}.

Our primary interest in previous Cepheid abundance studies has been the distribution of the 
elements in the galactic disc, and the motivation for this study of Ba abundances is no 
exception. While there is uncertainty about the abundances themselves, we shall persevere 
and consider the abundance distribution found from our Ba data.

\begin{figure}
\resizebox{\hsize}{!}           	
{\includegraphics {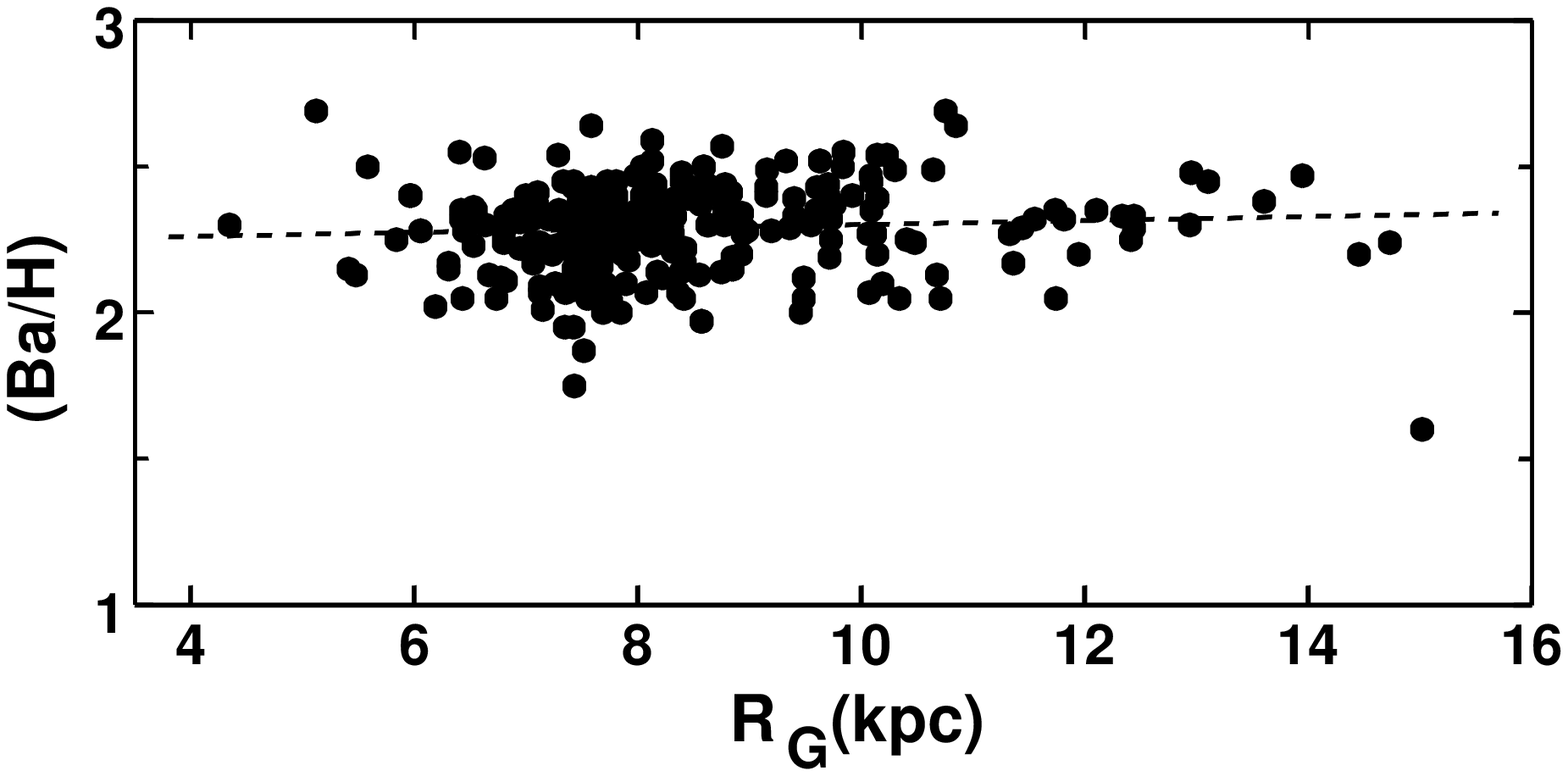}}
\resizebox{\hsize}{!}           	
{\includegraphics {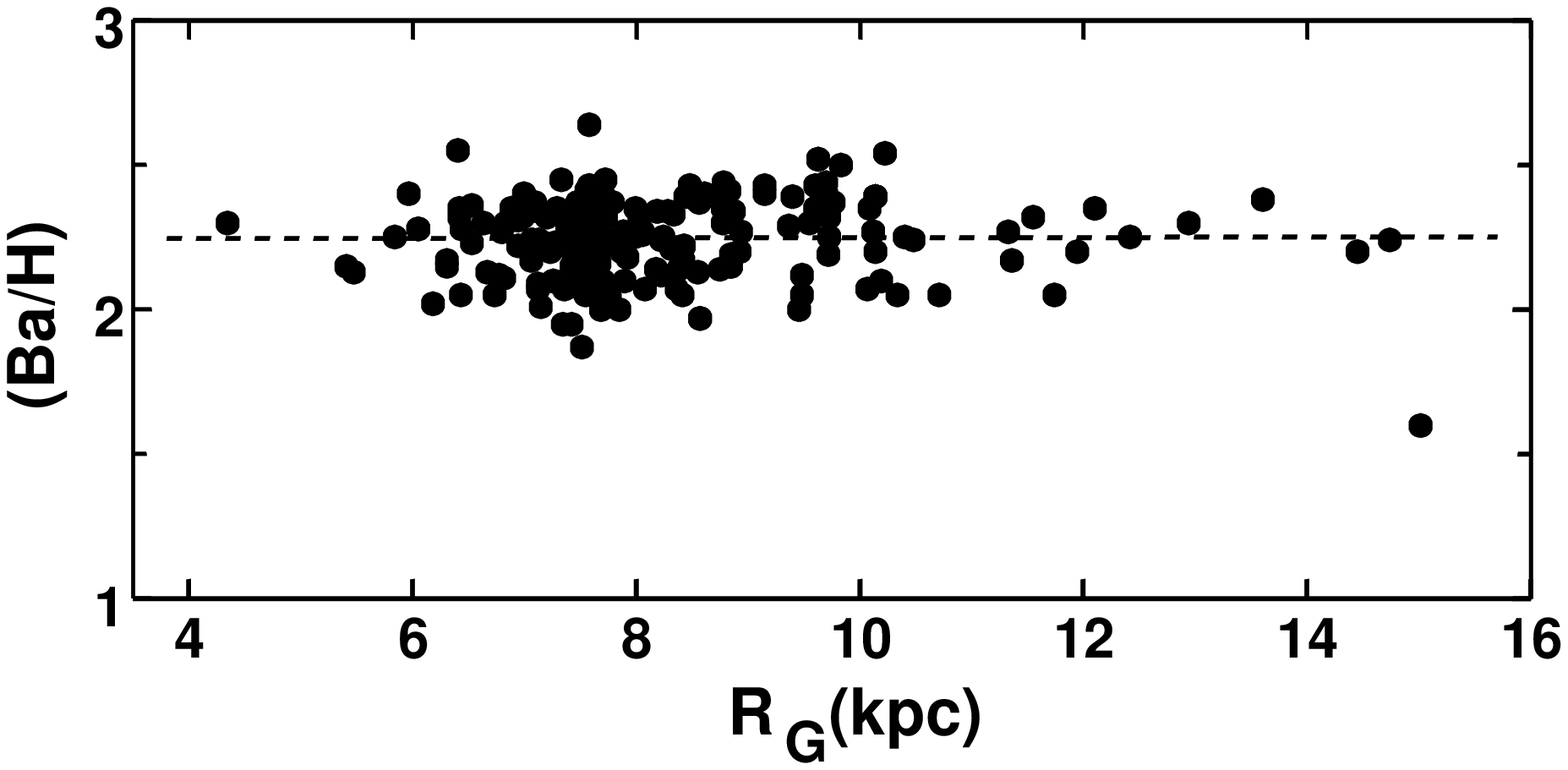}}
\caption[]{(Ba/H) data vs. galactocentric distance. Upper plot - the total sample of stars,
lower plot - part of the sample with V$_{\rm t} > 3.8$ km~s$^{-1}$.}
\label {Ba_R}
\end{figure}

In Figure \ref{Ba_R} we show the radial barium abundance distribution in galactic disc as derived from 
the total sample of the stars and from the sample with V$_{\rm t} > 3.8$ km~s$^{-1}$ (the distances are 
given in Table \ref{tabstars}). With the exception of a few deviant stars (such as EE Mon at 
R$_{\rm G} = 15$ kpc) the barium abundance distribution looks essentially flat. The mean value for the 
total sample is $<$(Ba/H)$>$ = 2.29 $\pm$ 0.15. If we use barium abundance data for the part of the sample 
with V$_{\rm t} > 3.8$ km~s$^{-1}$, the mean value is $<$(Ba/H)$>$ = 2.25 $\pm$ 0.13. 
Both values are the same within the estimated errors. What is important to note is that in both cases 
we have almost zero gradient.

\section{Discussion}

As a first approach to the Ba gradient, a linear regression was fitted to the data of Figures 
\ref{Ba_R}, with the formal result being: (Ba/H) = +0.0066 R$_{\rm G}$ + 2.2345 (the total sample), 
and (Ba/H) = +0.0003 R$_{\rm G}$ + 2.2456 (V$_{\rm t} > 3.8$ km~s$^{-1}$). From this, we conclude:

1) Barium abundances appear to be the same with a quite large scatter around the mean in the region 
spanning galactocentric distances from 4 to 15 kpc.

2) The mean (Ba/H) value for the solar neighborhood is near the solar barium abundance. 
The Cepheid mean values are 2.29 for the total sample and 2.25 for the stars with 
V$_{\rm t} > 3.8$ km~s$^{-1}$ versus a solar (NLTE) barium abundance of 2.17.

\begin{figure}
\resizebox{\hsize}{!}
{\includegraphics {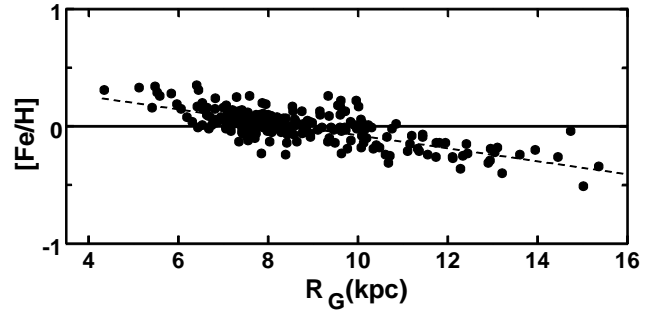}}
\caption[]{[Fe/H] vs. galactocentric distance from \citet{LL11} - Figure 1.}
\label {Fe_R}
\end{figure}

What could be the origin of a homogeneous distribution of barium in the galactic radius?   Let us 
consider the iron distribution in the disc presented in Figure \ref{Fe_R} (data from \citealt{Lu11}). 
This figure shows that the mean iron abundance in the range of galactocentric distances from 
9 to 14 kpc is about 0.2 dex lower than in the solar vicinity. This difference  would be about 0.3 dex  
if we adopt the metallicity distribution  with a  "step" at corotation as  proposed by \citet{Le11} 
based on a study of  open clusters. One could expect to detect lower barium abundance in the region 9-14 
kpc relative to the solar value, but this is not the case, as one can see in Figure \ref{Ba_R}.

If one adopts for lanthanum, praseodymium, neodymium, samarium, and europium 
the main $s$- and $r$-process dissection provided by \citet{Si04}, 
and plot their $r$-process fractions vs. abundance gradients derived for these 
elements by \citet{LL11}, then one can note the sligt (and quite 
loose) slope in the "abundance gradient - $r$-process fraction" dependence. 
Using this dependence together with barium $r$-proceess fraction 0.15, one can 
derive the expected barium abundence gradient of about --0.03 dex/kpc, which is in 
contrast to the zero slope from the present paper.

This discrepancy between observed and expected abundance gradient for
Ba abundance requires either a peculiar s-process pattern with lower 
"r"-process fraction for all above considered elements, or there is a 
systematic error in the Ba abundance analysis, an odd error that is not 
revealed by the temperature, gravity, or phase-dependent spectra. 

Another possibility could be connected to some unaccounted yet effect, which may 
be hidden in the literature La, Ce, Nd and Eu LTE abundance data derived for 
Cepheids, and respective gradients based on them. For instance, the NLTE 
deviations for these elemets may be different for the stars with (even slightly) 
different metallicities located at the different galactocentric distances.

Since there are no other literature large studies of the barium
abundance distributions in the disc, and thus there is no possibility 
to compare our barium data with the results of other specialists, we prefer 
to leave a detailed discussion of this problem for the future time,
until the new observational and theoretical data on barium and other 
heavy s-process element distributions in the galactic disc appear.
    
\section*{Acknowledgments}

SMA kindly acknowledges the Universidade de S\~ao Paulo (USP) and  
Instituto de Astronomia, Geof\'{\i}sica e Ci\^encias Atmosf\'ericas da
USP for support and hospitality during his stay in Brasil. 
SAK thanks the SCOPES grant No. IZ73Z0-128180/1 for partial financial support.

Authors thank anonymous referee for his/her comments.


\begin{table*}
\caption[]{Program stars, their phase parameters and barium abundance}
\label {tabstars}
\begin{tabular}{lccccccccc}
\hline\hline
Star & phase & T$_{\rm eff}$,\,K & $\log~g$ & V$_{\rm t}$, km~s$^{-1}$ &[Fe/H] &R$_{\rm G}$, (kpc) & 
$\rm \log \epsilon (Ba)_{NLTE}$ & $\sigma (Ba) $ & N lines  \\
\hline
\hline
AA Gem       &   0.625  &  5126 &   1.40  &  4.40  & -0.24 &  11.550  & 2.32  &  0.18 &    2    \\
AA Mon       &   0.106  &  6261 &   2.10  &  4.00  & -0.21 &  11.364  & 2.17  &  0.12 &    3    \\
AC Mon       &   0.081  &  6075 &   1.90  &  4.80  & -0.22 &  10.121  & 2.27  &  0.12 &    3    \\
AC Mon       &   0.919  &  6121 &   2.20  &  5.80  & -0.07 &  10.121  & 2.27  &  0.12 &    3    \\
AD Cru       &   0.030  &  6000 &   2.00  &  4.00  &  0.06 &  6.987   & 2.32  &  0.13 &    3    \\
AD Gem       &   0.818  &  5784 &   2.15  &  4.30  & -0.19 &  10.481  & 2.24  &  0.18 &    2    \\
AE Tau       &   0.377  &  5981 &   2.30  &  4.00  & -0.19 &  11.326  & 2.27  &  0.13 &    3    \\
AE Vel       &   0.732  &  5460 &   1.70  &  4.50  &  0.05 &  7.983   & 2.25  &  0.20 &    3    \\
AG Cru       &   0.039  &  6628 &   2.20  &  4.70  & -0.13 &  7.346   & 1.95  &  0.13 &    3    \\
AH Vel       &   0.383  &  6037 &   2.20  &  4.30  &  0.10 &  7.996   & 2.35  &  0.13 &    3    \\
AO Aur       &   0.707  &  5459 &   1.70  &  3.60  & -0.14 &  11.813  & 2.32  &  0.18 &    2    \\
AP Pup       &   0.052  &  6233 &   2.10  &  4.20  &  0.06 &  8.189   & 2.34  &  0.13 &    3    \\
AP Sgr       &   0.093  &  6328 &   2.02  &  3.46  &  0.10 &  7.101   & 2.41  &  0.15 &    6    \\
AQ Car       &   0.998  &  5815 &   1.90  &  4.90  &  0.06 &  7.631   & 2.22  &  0.13 &    3    \\
AQ Pup       &   0.756  &  4937 &   0.60  &  5.50  & -0.14 &  9.487   & 2.12  &  0.13 &    3    \\
AQ Pup       &   0.798  &  4958 &   0.60  &  5.70  & -0.14 &  9.487   & 2.05  &  0.15 &    3    \\
AS Per       &   0.580  &  5550 &   1.72  &  3.35  &  0.10 &  9.154   & 2.49  &  0.15 &    5    \\
AT Pup       &   0.658  &  6455 &   2.10  &  5.00  & -0.14 &  8.412   & 2.05  &  0.15 &    3    \\
AV Cir       &   0.207  &  6169 &   2.10  &  3.30  &  0.10 &  7.516   & 2.30  &  0.13 &    3    \\
AV Sgr       &   0.982  &  5875 &   1.75  &  4.90  &  0.34 &  5.475   & 2.13  &  0.15 &    3    \\
AW Per       &   0.295  &  5989 &   1.90  &  3.60  &  0.00 &  8.622   & 2.30  &  0.13 &    5    \\
AW Per       &   0.451  &  5836 &   2.00  &  3.70  &  0.00 &  8.622   & 2.30  &  0.13 &    6    \\
AX Cir       &   0.119  &  5761 &   1.80  &  3.70  & -0.05 &  7.530   & 2.16  &  0.13 &    3    \\
AX Cir       &   0.745  &  5648 &   1.90  &  4.50  & -0.07 &  7.530   & 2.20  &  0.12 &    3    \\
AY Cen       &   0.999  &  5933 &   2.10  &  4.50  &  0.01 &  7.424   & 2.25  &  0.10 &    3    \\
AZ Cen       &   0.947  &  6425 &   2.30  &  4.00  & -0.05 &  7.413   & 2.21  &  0.10 &    3    \\
BB Cen       &   0.183  &  6202 &   2.30  &  4.60  &  0.13 &  7.126   & 2.24  &  0.13 &    3    \\
BB Gem       &   0.589  &  6035 &   2.40  &  3.00  & -0.09 &  10.641  & 2.49  &  0.18 &    2    \\
BB Her       &   0.259  &  5750 &   1.80  &  4.20  &  0.16 &  6.054   & 2.28  &  0.13 &    6    \\
BB Her       &   0.389  &  5556 &   1.70  &  4.20  &  0.13 &  6.054   & 2.28  &  0.14 &    6    \\
BB Sgr       &   0.878  &  5859 &   1.87  &  4.47  &  0.08 &  7.138   & 2.34  &  0.17 &    5    \\
BC Pup       &   0.062  &  6695 &   2.00  &  3.50  & -0.20 &  12.440  & 2.29  &  0.13 &    3    \\
BC Pup       &   0.341  &  5879 &   1.90  &  3.30  & -0.24 &  12.440  & 2.33  &  0.14 &    3    \\
BD Cas       &   0.772  &  6200 &   2.50  &  5.00  & -0.07 &  8.574   & 1.97  &  0.18 &    3    \\
BD Cas       &   0.073  &  6075 &   2.30  &  4.50  & -0.07 &  8.574   & 1.97  &  0.17 &    3    \\
$\beta$ Dor  &   0.128  &  5618 &   1.60  &  4.30  & -0.01 &  7.896   & 2.10  &  0.13 &    6    \\
BF Oph       &   0.033  &  6246 &   2.30  &  4.60  &  0.05 &  7.121   & 2.09  &  0.10 &    3    \\
BF Oph       &          &  5541 &   1.90  &  4.10  & -0.01 &  7.121   & 2.07  &  0.11 &    3    \\
BG Cru       &   0.150  &  6309 &   2.30  &  4.20  &  0.06 &  7.694   & 2.25  &  0.16 &    3    \\
BG Cru       &   0.188  &  6101 &   2.00  &  3.80  & -0.02 &  7.694   & 2.25  &  0.15 &    6    \\
BG Lac       &   0.147  &  5923 &   1.90  &  3.80  &  0.01 &  8.158   & 2.44  &  0.13 &    5    \\
BG Lac       &   0.332  &  5625 &   1.85  &  3.60  &  0.02 &  8.158   & 2.44  &  0.13 &    6    \\
BG Vel       &   0.849  &  5727 &   1.90  &  5.30  & -0.01 &  7.922   & 2.18  &  0.13 &    3    \\
BM Per       &   0.344  &  5313 &   0.54  &  3.40  &  0.10 &  10.851  & 2.64  &  0.18 &    5    \\
BN Pup       &   0.396  &  5105 &   1.00  &  3.50  &  0.01 &  9.920   & 2.40  &  0.15 &    3    \\
BP Cir       &   0.033  &  6533 &   2.40  &  3.70  & -0.06 &  7.431   & 1.75  &  0.15 &    3    \\
CD Cyg       &   0.376  &  5211 &   1.00  &  3.50  &  0.10 &  7.474   & 2.40  &  0.13 &    6    \\
CD Cyg       &   0.434  &  5108 &   1.10  &  4.00  &  0.15 &  7.474   & 2.37  &  0.14 &    6    \\
CD Cyg       &   0.519  &  4979 &   1.00  &  4.00  &  0.08 &  7.474   & 2.37  &  0.13 &    6    \\
CE Cas A     &   0.732  &  5817 &   1.73  &  3.81  &  0.18 &  9.549   & 2.30  &  0.20 &    5    \\
CE Pup       &   0.804  &  5846 &   1.30  &  5.80  & -0.04 &  14.739  & 2.24  &  0.13 &    3    \\
CF Cas       &   0.584  &  5454 &   1.70  &  4.30  &  0.01 &  9.702   & 2.40  &  0.15 &    6    \\
CF Cas       &   0.980  &  6115 &   2.00  &  4.00  & -0.03 &  9.702   & 2.44  &  0.16 &    6    \\
CF Cas       &   0.238  &  5704 &   1.90  &  3.70  &  0.02 &  9.702   & 2.40  &  0.15 &    6    \\
CH Cas       &   0.108  &  6319 &   1.80  &  5.50  & -0.08 &  9.601   & 2.43  &  0.17 &    5    \\
CN Car       &   0.066  &  6331 &   2.20  &  4.30  &  0.06 &  7.802   & 2.22  &  0.13 &    3    \\
CP Cep       &   0.243  &  5150 &   1.30  &  4.10  & -0.14 &  9.600   & 2.35  &  0.17 &    3    \\
CR Cep       &   0.634  &  5443 &   1.33  &  4.02  & -0.06 &  8.442   & 2.39  &  0.15 &    6    \\
CS Ori       &   0.679  &  6539 &   2.50  &  3.60  & -0.26 &  11.738  & 2.35  &  0.15 &    3    \\
CU Mon       &   0.655  &  5677 &   1.90  &  4.50  & -0.25 &  14.451  & 2.20  &  0.15 &    3    \\
CV Mon       &   0.178  &  5897 &   2.00  &  4.10  & -0.03 &  9.461   & 2.00  &  0.15 &    3    \\
CX Vel       &   0.003  &  6251 &   2.20  &  5.30  &  0.06 &  8.360   & 2.07  &  0.13 &    3    \\
CY Car       &   0.095  &  6042 &   2.20  &  4.00  &  0.10 &  7.471   & 2.32  &  0.13 &    3    \\
CY Cas       &   0.391  &  5369 &   0.85  &  3.45  &  0.06 &  10.062  & 2.27  &  0.18 &    5    \\
\hline                         
\end{tabular}                  
\end{table*}                    

\begin{table*}
\contcaption{}
\begin{tabular}{lccccccccc}
\hline
\hline
Star & phase & T$_{\rm eff}$,\,K & $\log~g$ & V$_{\rm t}$, km~s$^{-1}$ &[Fe/H] &R$_{\rm G}$, (kpc) & 
$\rm \log \epsilon (Ba)_{NLTE}$ & $\sigma (Ba) $ & N lines  \\
\hline
DD Cas       &   0.039  &  5901 &   1.80  &  4.50  &  0.04 &  9.602   & 2.35  &  0.18 &    5    \\
DF Cas       &   0.401  &  5644 &   2.20  &  4.65  &  0.13 &  9.718   & 2.19  &  0.18 &    3    \\
DK Vel       &   0.009  &  6448 &   2.20  &  3.70  & -0.02 &  8.126   & 2.23  &  0.13 &    3    \\
DL Cas       &   0.105  &  5860 &   1.70  &  4.70  & -0.01 &  8.846   & 2.15  &  0.13 &    6    \\
DL Cas       &   0.229  &  5786 &   1.70  &  4.20  & -0.01 &  8.846   & 2.19  &  0.13 &    6    \\
DL Cas       &   0.473  &  5438 &   1.40  &  4.00  & -0.01 &  8.846   & 2.15  &  0.14 &    6    \\
DR Vel       &   0.165  &  5482 &   1.50  &  3.80  &  0.08 &  8.036   & 2.40  &  0.13 &    3    \\
DT Cyg       &   0.651  &  6406 &   2.60  &  3.70  &  0.16 &  7.805   & 2.37  &  0.13 &    6    \\
DT Cyg       &   0.705  &  6286 &   2.35  &  3.10  &  0.10 &  7.805   & 2.45  &  0.13 &    6    \\
DT Cyg       &   0.769  &  6339 &   2.45  &  3.40  &  0.13 &  7.805   & 2.40  &  0.13 &    6    \\
DX Gem       &   0.353  &  6096 &   1.90  &  2.80  & -0.02 &  10.755  & 2.69  &  0.20 &    2    \\
DY Car       &   0.956  &  6533 &   2.30  &  5.30  & -0.07 &  7.686   & 2.00  &  0.13 &    3    \\
EE Mon       &   0.209  &  6134 &   1.80  &  4.50  & -0.51 &  15.018  & 1.60  &  0.18 &    3    \\
EK Mon       &   0.918  &  6001 &   2.00  &  4.50  & -0.10 &  10.191  & 2.10  &  0.20 &    3    \\
ER Car       &   0.872  &  5874 &   2.10  &  6.00  &  0.03 &  7.597   & 2.07  &  0.13 &    3    \\
$\eta$ Aql   &   0.263  &  5800 &   1.90  &  4.00  &  0.09 &  7.713   & 2.35  &  0.13 &    6    \\
$\eta$ Aql   &   0.404  &  5612 &   1.75  &  3.90  &  0.10 &  7.713   & 2.39  &  0.13 &    6    \\
$\eta$ Aql   &   0.462  &  5508 &   1.80  &  4.00  &  0.08 &  7.713   & 2.39  &  0.13 &    6    \\
EU Tau       &   0.172  &  6203 &   2.00  &  3.00  & -0.06 &  8.933   & 2.30  &  0.13 &    6    \\
EU Tau       &   0.641  &  6014 &   2.20  &  3.30  & -0.06 &  8.933   & 2.34  &  0.13 &    6    \\
EW Sct       &   0.572  &  5655 &   1.70  &  3.40  &  0.04 &  7.568   & 2.36  &  0.13 &    6    \\
EW Sct       &   0.722  &  5728 &   1.80  &  3.50  &  0.04 &  7.568   & 2.36  &  0.13 &    6    \\
EX Vel       &   0.954  &  5902 &   2.00  &  5.00  &  0.05 &  8.872   & 2.34  &  0.13 &    3    \\
FF Aql       &   0.219  &  6172 &   2.15  &  4.30  &  0.00 &  7.629   & 2.32  &  0.18 &    6    \\
FF Aql       &   0.846  &  6328 &   2.10  &  5.20  &  0.09 &  7.629   & 2.32  &  0.17 &    6    \\
FF Aql       &   0.995  &  6425 &   2.10  &  4.90  &  0.02 &  7.629   & 2.22  &  0.18 &    6    \\
FG Mon       &   0.523  &  5626 &   1.60  &  3.00  & -0.20 &  13.944  & 2.47  &  0.15 &    3    \\
FG Vel       &   0.756  &  5411 &   1.50  &  4.40  & -0.05 &  8.286   & 2.34  &  0.18 &    3    \\
FI Car       &   0.215  &  5215 &   1.20  &  3.80  & -0.01 &  8.107   & 2.29  &  0.13 &    6    \\
FI Mon       &   0.356  &  5938 &   2.00  &  3.50  & -0.18 &  13.106  & 2.45  &  0.15 &    6    \\
FM Aql       &   0.432  &  5590 &   1.60  &  3.70  &  0.06 &  7.290   & 2.54  &  0.15 &    6    \\
FM Cas       &   0.669  &  5395 &   1.60  &  4.20  & -0.12 &  8.941   & 2.27  &  0.15 &    5    \\
FN Vel       &   0.221  &  5809 &   1.90  &  3.70  &  0.06 &  7.850   & 2.32  &  0.15 &    3    \\
FR Car       &   0.011  &  5905 &   2.00  &  5.00  &  0.02 &  7.387   & 2.33  &  0.13 &    6    \\
GH Car       &   0.965  &  6336 &   2.00  &  5.50  & -0.01 &  7.414   & 2.27  &  0.13 &    3    \\
GH Lup       &   0.142  &  5460 &   1.60  &  4.60  &  0.08 &  6.944   & 2.22  &  0.13 &    3    \\
GQ Ori       &   0.074  &  6002 &   1.80  &  4.10  &  0.01 &  10.226  & 2.54  &  0.18 &    5    \\
GU Nor       &   0.145  &  6036 &   2.20  &  3.60  &  0.15 &  6.553   & 2.35  &  0.13 &    3    \\
GX Car       &   0.996  &  6297 &   2.20  &  5.30  &  0.01 &  7.762   & 2.04  &  0.13 &    3    \\
HW Car       &   0.947  &  5671 &   2.00  &  5.20  &  0.04 &  7.556   & 2.24  &  0.13 &    3    \\
HW Pup       &   0.432  &  5869 &   1.55  &  3.60  & -0.15 &  12.332  & 2.33  &  0.15 &    3    \\
IO Car       &   0.072  &  6145 &   1.80  &  6.00  & -0.05 &  8.078   & 2.07  &  0.15 &    3    \\
IR Cep       &   0.385  &  6104 &   2.10  &  2.90  &  0.11 &  8.037   & 2.50  &  0.15 &    6    \\
IT Car       &   0.950  &  5794 &   1.80  &  4.70  &  0.06 &  7.451   & 2.17  &  0.13 &    3    \\
KK Cen       &   0.012  &  5894 &   1.90  &  5.50  &  0.12 &  7.542   & 2.32  &  0.13 &    3    \\
KN Cen       &   0.488  &  4870 &   1.70  &  4.50  &  0.35 &  6.404   & 2.55  &  0.18 &    3    \\
KQ Sco       &   0.940  &  5058 &   1.10  &  5.70  &  0.16 &  5.409   & 2.15  &  0.17 &    3    \\
L Car        &   0.094  &  5298 &   1.10  &  4.50  &  0.05 &  7.794   & 2.37  &  0.15 &    3    \\
LR TrA       &   0.086  &  5943 &   2.20  &  4.40  &  0.25 &  7.291   & 2.35  &  0.13 &    3    \\
MM Per       &   0.358  &  5746 &   2.15  &  3.40  & -0.01 &  10.304  & 2.49  &  0.18 &    2    \\
MW Cyg       &   0.458  &  5522 &   1.65  &  4.00  &  0.05 &  7.553   & 2.24  &  0.15 &    5    \\
MY Pup       &   0.114  &  6317 &   2.20  &  4.80  & -0.14 &  8.028   & 2.32  &  0.13 &    3    \\
MZ Cen       &   0.207  &  5816 &   1.90  &  5.20  &  0.20 &  6.531   & 2.23  &  0.15 &    3    \\
NT Pup       &   0.219  &  5561 &   1.20  &  4.00  & -0.15 &  12.416  & 2.25  &  0.13 &    3    \\
QY Cen       &   0.004  &  6205 &   2.00  &  6.30  &  0.16 &  6.638   & 2.30  &  0.20 &    1    \\
R Cru        &   0.100  &  6203 &   2.00  &  4.00  &  0.08 &  7.539   & 2.25  &  0.13 &    3    \\
R Mus        &   0.184  &  5985 &   2.00  &  4.00  &  0.10 &  7.502   & 2.32  &  0.13 &    3    \\
R TrA        &   0.167  &  6121 &   2.20  &  3.80  &  0.06 &  7.475   & 2.22  &  0.13 &    3    \\
RR Lac       &   0.518  &  5646 &   1.80  &  4.10  &  0.00 &  8.554   & 2.13  &  0.15 &    5    \\
RS Ori       &   0.831  &  5934 &   1.85  &  5.10  & -0.02 &  9.363   & 2.29  &  0.15 &    3    \\
RS Pup       &   0.905  &  5068 &   1.00  &  5.00  &  0.20 &  8.540   & 2.38  &  0.13 &    3    \\
RS Pup       &   0.935  &  5073 &   1.00  &  4.70  &  0.16 &  8.540   & 2.38  &  0.13 &    3    \\
RT Aur       &   0.535  &  5696 &   2.15  &  3.30  &  0.05 &  8.304   & 2.30  &  0.13 &    6    \\
RT Aur       &   0.583  &  5686 &   1.85  &  3.40  &  0.07 &  8.304   & 2.28  &  0.14 &    6    \\
\hline                         
\end{tabular}                  
\end{table*}                    

\begin{table*}
\contcaption{}
\begin{tabular}{lccccccccc}
\hline
\hline
Star & phase & T$_{\rm eff}$,\,K & $\log~g$ & V$_{\rm t}$, km~s$^{-1}$ &[Fe/H] &R$_{\rm G}$, (kpc) & 
$\rm \log \epsilon (Ba)_{NLTE}$ & $\sigma (Ba) $ & N lines  \\
\hline
RT Mus       &   0.204  &  6236 &   2.40  &  4.00  &  0.02 &  7.426   & 2.08  &  0.13 &    3    \\
RV Sco       &   0.914  &  6172 &   2.30  &  5.40  &  0.03 &  6.734   & 2.05  &  0.13 &    3    \\
RX Aur       &   0.135  &  6165 &   2.00  &  4.70  &  0.04 &  9.397   & 2.39  &  0.15 &    3    \\
RX Aur       &   0.334  &  5677 &   1.40  &  3.70  & -0.06 &  9.397   & 2.33  &  0.14 &    3    \\
RX Cam       &   0.212  &  5971 &   1.95  &  4.10  &  0.07 &  8.614   & 2.40  &  0.15 &    6    \\
RX Cam       &   0.461  &  5485 &   1.60  &  3.50  &  0.05 &  8.614   & 2.44  &  0.14 &    6    \\
RY CMa       &   0.897  &  6096 &   2.10  &  4.30  &  0.02 &  8.781   & 2.44  &  0.20 &    2    \\
RY Sco       &   0.746  &  5519 &   1.50  &  6.30  &  0.04 &  6.668   & 2.13  &  0.20 &    6    \\
RY Vel       &   0.412  &  5466 &   1.30  &  4.60  &  0.10 &  7.731   & 2.32  &  0.13 &    3    \\
RZ Gem       &   0.325  &  5906 &   2.20  &  3.10  & -0.12 &  9.838   & 2.55  &  0.20 &    2    \\
RZ Vel       &   0.037  &  6669 &   1.55  &  6.50  & -0.03 &  8.225   & 2.24  &  0.15 &    3    \\
RZ Vel       &   0.988  &  6549 &   1.65  &  6.80  & -0.11 &  8.225   & 2.12  &  0.14 &    3    \\
S Cru        &   0.983  &  6464 &   2.10  &  4.10  & -0.12 &  7.553   & 2.20  &  0.13 &    3    \\
S Mus        &   0.106  &  5752 &   1.80  &  4.40  & -0.02 &  7.564   & 2.34  &  0.13 &    3    \\
S Nor        &   0.124  &  5859 &   2.00  &  5.00  &  0.07 &  7.169   & 2.23  &  0.13 &    3    \\
S Sge        &   0.295  &  5887 &   1.80  &  3.90  &  0.10 &  7.552   & 2.41  &  0.18 &    3    \\
S Sge        &   0.345  &  5661 &   1.90  &  4.40  &  0.10 &  7.552   & 2.30  &  0.17 &    3    \\
S Sge        &   0.463  &  5476 &   1.60  &  3.90  &  0.10 &  7.552   & 2.30  &  0.16 &    3    \\
S Sge        &   0.502  &  5412 &   1.55  &  3.75  &  0.10 &  7.552   & 2.37  &  0.18 &    3    \\
S Sge        &   0.820  &  5799 &   2.10  &  5.50  &  0.10 &  7.552   & 2.25  &  0.19 &    3    \\
S Sge        &   0.911  &  6115 &   2.00  &  5.00  &  0.07 &  7.552   & 2.34  &  0.17 &    3    \\
S Sge        &   0.935  &  6198 &   2.10  &  5.00  &  0.08 &  7.552   & 2.21  &  0.16 &    3    \\
S Sge        &   0.940  &  6239 &   2.20  &  4.10  &  0.10 &  7.552   & 2.25  &  0.17 &    3    \\
S TrA        &   0.176  &  5976 &   2.10  &  4.20  &  0.12 &  7.283   & 2.22  &  0.13 &    3    \\
S Vul        &   0.516  &  5166 &   0.50  &  6.40  & -0.04 &  7.067   & 2.24  &  0.18 &    6    \\
S Vul        &   0.557  &  5123 &   0.50  &  6.60  & -0.01 &  7.067   & 2.24  &  0.17 &    6    \\
ST Tau       &   0.132  &  6268 &   2.00  &  3.50  &       &  8.834   & 2.41  &  0.15 &    6    \\
ST Tau       &   0.869  &  6519 &   2.50  &  4.40  &       &  8.834   & 2.41  &  0.16 &    6    \\
ST Vel       &   0.978  &  6255 &   2.10  &  4.30  &  0.00 &  8.183   & 2.34  &  0.13 &    3    \\
SU Cas       &   0.405  &  6162 &   2.35  &  3.00  &  0.08 &  8.127   & 2.40  &  0.15 &    6    \\
SU Cas       &   0.616  &  6112 &   2.30  &  2.80  &  0.06 &  8.127   & 2.59  &  0.14 &    6    \\
SU Cas       &   0.855  &  6520 &   2.60  &  3.30  &  0.06 &  8.127   & 2.40  &  0.15 &    6    \\
SU Cas       &   0.926  &  6603 &   2.80  &  3.50  &  0.02 &  8.127   & 2.31  &  0.15 &    6    \\
SU Cyg       &   0.415  &  5956 &   2.10  &  3.20  &  0.00 &  7.603   & 2.39  &  0.17 &    6    \\
SU Cyg       &   0.917  &  6197 &   2.20  &  4.25  & -0.03 &  7.603   & 2.32  &  0.18 &    6    \\
SU Cyg       &   0.940  &  6314 &   2.40  &  4.50  & -0.03 &  7.603   & 2.32  &  0.18 &    6    \\
SV Mon       &   0.047  &  6141 &   1.50  &  4.60  & -0.05 &  10.143  & 2.39  &  0.18 &    5    \\
SV Mon       &   0.223  &  5464 &   1.10  &  4.20  &  0.06 &  10.143  & 2.39  &  0.15 &    6    \\
SV Mon       &   0.352  &  5204 &   0.90  &  3.30  & -0.01 &  10.143  & 2.54  &  0.17 &    6    \\
SV Mon       &   0.559  &  4924 &   1.10  &  4.50  &  0.00 &  10.143  & 2.20  &  0.17 &    3    \\
SV Mon       &   0.784  &  5263 &   1.35  &  7.00  & -0.07 &  10.143  & 2.39  &  0.18 &    6    \\
SV Mon       &   0.851  &  5522 &   1.50  &  6.00  & -0.05 &  10.143  & 2.39  &  0.17 &    6    \\
SV Mon       &   0.916  &  5482 &   1.40  &  4.90  & -0.04 &  10.143  & 2.39  &  0.18 &    6    \\
SV Per       &   0.894  &  5734 &   1.70  &  4.30  & -0.05 &  10.088  & 2.35  &  0.15 &    5    \\
SV Vel       &   0.939  &  6064 &   2.00  &  6.00  &  0.08 &  7.594   & 2.07  &  0.15 &    3    \\
SV Vul       &   0.070  &  5856 &   1.20  &  5.90  &  0.11 &  7.261   & 2.10  &  0.13 &    6    \\
SV Vul       &   0.414  &  5005 &   0.65  &  5.00  & -0.01 &  7.261   & 2.10  &  0.14 &    6    \\
SW Cas       &   0.791  &  5749 &   2.00  &  5.10  &  0.02 &  8.747   & 2.14  &  0.20 &    2    \\
SW Vel       &   0.348  &  5010 &   1.00  &  4.40  &  0.00 &  8.427   & 2.22  &  0.15 &    3    \\
SX Car       &   0.943  &  6513 &   2.10  &  4.50  & -0.09 &  7.586   & 2.17  &  0.13 &    3    \\
SX Vel       &   0.852  &  6280 &   2.00  &  4.50  &  0.00 &  8.245   & 2.25  &  0.13 &    3    \\
SY Cas       &   0.985  &  5718 &   2.00  &  4.30  & -0.06 &  8.928   & 2.20  &  0.15 &    6    \\
SY Nor       &   0.164  &  5641 &   1.80  &  4.70  &  0.31 &  6.442   & 2.28  &  0.20 &    3    \\
SZ Aql       &   0.015  &  6559 &   2.00  &  6.00  &  0.13 &  6.421   & 2.32  &  0.13 &    6    \\
SZ Aql       &   0.402  &  5147 &   1.10  &  3.85  &  0.25 &  6.421   & 2.35  &  0.14 &    6    \\
SZ Cas       &   0.694  &  6157 &   1.60  &  4.30  &  0.02 &  9.832   & 2.50  &  0.20 &    5    \\
SZ Cyg       &   0.293  &  5109 &   1.30  &  3.90  &  0.06 &  8.059   & 2.27  &  0.15 &    6    \\
SZ Tau       &   0.296  &  6037 &   2.20  &  3.60  &  0.08 &  8.394   & 2.45  &  0.13 &    6    \\
SZ Tau       &   0.426  &  5867 &   2.00  &  3.30  &  0.08 &  8.394   & 2.48  &  0.13 &    6    \\
T Ant        &   0.173  &  6265 &   2.10  &  4.40  & -0.24 &  8.384   & 2.14  &  0.12 &    2    \\
T Cru        &   0.958  &  5940 &   2.20  &  5.10  &  0.09 &  7.546   & 2.25  &  0.13 &    3    \\
\hline                         
\end{tabular}                  
\end{table*}                    

\begin{table*}
\contcaption{}
\begin{tabular}{lccccccccc}
\hline
\hline
Star & phase & T$_{\rm eff}$,\,K & $\log~g$ & V$_{\rm t}$, km~s$^{-1}$ &[Fe/H] &R$_{\rm G}$, (kpc) & 
$\rm \log \epsilon (Ba)_{NLTE}$ & $\sigma (Ba) $ & N lines  \\
\hline
T Mon        &   0.562  &  4853 &   0.80  &  4.20  &  0.21 &  9.150   & 2.43  &  0.13 &    3    \\
T Mon        &   0.600  &  4828 &   1.10  &  4.70  &  0.11 &  9.150   & 2.40  &  0.14 &    3    \\
T Vel        &   0.827  &  5692 &   2.00  &  4.80  & -0.02 &  8.054   & 2.26  &  0.15 &    3    \\
T Vul        &   0.201  &  6077 &   2.00  &  3.55  &  0.02 &  7.760   & 2.35  &  0.18 &    6    \\
T Vul        &   0.302  &  5899 &   2.00  &  3.30  &  0.03 &  7.760   & 2.43  &  0.17 &    6    \\
TT Aql       &   0.457  &  5080 &   1.10  &  3.60  &  0.12 &  7.099   & 2.37  &  0.13 &    6    \\
TT Aql       &   0.920  &  5630 &   1.60  &  5.10  &  0.09 &  7.099   & 2.37  &  0.14 &    6    \\
TW CMa       &   0.059  &  6142 &   1.80  &  3.90  & -0.18 &  9.764   & 2.37  &  0.20 &    2    \\
TW Mon       &   0.167  &  5770 &   1.70  &  4.00  & -0.24 &  13.608  & 2.38  &  0.13 &    3    \\
TW Nor       &   0.094  &  5979 &   2.00  &  5.60  &  0.28 &  5.843   & 2.25  &  0.15 &    3    \\
TX Cyg       &   0.936  &  6054 &   2.00  &  5.60  &  0.08 &  7.869   & 2.20  &  0.15 &    6    \\
TX Del       &   0.005  &  6217 &   1.80  &  6.00  &  0.24 &  6.820   & 2.11  &  0.15 &    3    \\
TX Mon       &   0.122  &  5871 &   1.80  &  4.90  & -0.14 &  11.742  & 2.05  &  0.18 &    3    \\
TZ Mon       &   0.817  &  5008 &   2.20  &  1.50  & -0.03 &  11.439  & 2.29  &  0.13 &    3    \\
TZ Mus       &   0.024  &  6348 &   2.10  &  4.60  & -0.01 &  7.063   & 2.17  &  0.15 &    3    \\
U Car        &   0.968  &  4934 &   0.80  &  5.80  &  0.01 &  7.537   & 2.27  &  0.15 &    3    \\
U Nor        &   0.476  &  5426 &   1.40  &  3.60  &  0.15 &  6.815   & 2.33  &  0.10 &    3    \\
U Vul        &   0.290  &  5965 &   1.85  &  4.10  &  0.13 &  7.584   & 2.43  &  0.13 &    6    \\
U Vul        &   0.415  &  5655 &   1.75  &  4.00  &  0.10 &  7.584   & 2.43  &  0.14 &    6    \\
UU Mus       &   0.041  &  6175 &   1.80  &  5.40  &  0.05 &  7.054   & 2.34  &  0.15 &    3    \\
UW Car       &   0.035  &  6618 &   2.10  &  4.30  & -0.06 &  7.617   & 2.05  &  0.12 &    3    \\
UX Car       &   0.902  &  6442 &   2.60  &  4.70  &  0.02 &  7.663   & 2.19  &  0.12 &    3    \\
UY Car       &   0.946  &  6376 &   2.40  &  5.70  &  0.03 &  7.548   & 2.05  &  0.15 &    3    \\
UY Mon       &   0.674  &  6181 &   2.20  &  2.70  & -0.08 &  10.085  & 2.47  &  0.18 &    2    \\
UZ Car       &   0.044  &  6079 &   2.00  &  4.30  &  0.07 &  7.544   & 2.31  &  0.15 &    2    \\
UZ Sct       &   0.455  &  5127 &   1.50  &  3.30  &  0.33 &  5.125   & 2.69  &  0.15 &    3    \\
V Car        &   0.889  &  5906 &   2.10  &  5.20  &  0.01 &  7.877   & 2.22  &  0.12 &    2    \\
V Cen        &   0.236  &  5705 &   2.10  &  3.90  &  0.01 &  7.426   & 2.15  &  0.12 &    3    \\
V Cen        &   0.781  &  5616 &   1.80  &  4.80  &  0.01 &  7.426   & 1.95  &  0.12 &    3    \\
V Cen        &   0.952  &  6427 &   2.30  &  4.60  &  0.01 &  7.426   & 2.15  &  0.13 &    3    \\
V Lac        &   0.098  &  6470 &   1.90  &  3.60  & -0.05 &  8.485   & 2.40  &  0.18 &    5    \\
V Vel        &   0.043  &  6364 &   2.10  &  5.00  & -0.23 &  7.849   & 2.00  &  0.13 &    3    \\
V335 Aur     &   0.622  &  5750 &   2.20  &  3.50  & -0.27 &  9.193   & 2.28  &  0.20 &    1    \\
V339 Cen     &   0.142  &  5923 &   2.00  &  5.30  &  0.04 &  6.774   & 2.12  &  0.13 &    3    \\
V340 Ara     &   0.167  &  5472 &   1.50  &  4.00  &  0.31 &  4.344   & 2.30  &  0.15 &    3    \\
V340 Nor     &   0.193  &  5595 &   1.75  &  4.50  &  0.08 &  6.306   & 2.15  &  0.13 &    3    \\
V340 Nor     &   0.430  &  5733 &   2.00  &  5.30  &  0.08 &  6.306   & 2.17  &  0.14 &    3    \\
V350 Sgr     &   0.468  &  5557 &   1.90  &  3.70  &  0.15 &  7.065   & 2.34  &  0.15 &    6    \\
V351 Cep     &   0.301  &  6005 &   2.10  &  3.30  &  0.03 &  8.313   & 2.27  &  0.18 &    3    \\
V351 Cep     &   0.425  &  5944 &   2.50  &  4.30  &  0.03 &  8.313   & 2.21  &  0.18 &    2    \\
V367 Sct     &          &  5891 &   2.10  &  4.25  & -0.01 &  6.431   & 2.05  &  0.20 &    3    \\
V378 Cen     &   0.131  &  6184 &   2.00  &  4.70  & -0.02 &  7.054   & 2.35  &  0.13 &    3    \\
V379 Cas     &   0.515  &  5943 &   2.00  &  3.50  &  0.04 &  8.592   & 2.50  &  0.15 &    6    \\
V379 Cas     &   0.740  &  5969 &   2.00  &  3.50  &  0.01 &  8.592   & 2.50  &  0.14 &    6    \\
V381 Cen     &   0.896  &  6224 &   2.30  &  5.10  &  0.02 &  7.236   & 2.20  &  0.13 &    3    \\
V386 Cyg     &   0.961  &  6284 &   2.20  &  4.60  & -0.04 &  7.888   & 2.27  &  0.15 &    6    \\
V397 Car     &   0.903  &  6036 &   2.30  &  4.80  &  0.03 &  7.674   & 2.16  &  0.13 &    3    \\
V402 Cyg     &   0.514  &  5518 &   1.55  &  3.70  & -0.06 &  7.600   & 2.27  &  0.18 &    6    \\
V419 Cen     &   0.905  &  6276 &   2.10  &  5.60  &  0.07 &  7.411   & 2.27  &  0.13 &    3    \\
V440 Per     &   0.275  &  6048 &   2.20  &  4.80  & -0.02 &  8.478   & 2.43  &  0.18 &    6    \\
V440 Per     &   0.807  &  6041 &   2.00  &  4.80  & -0.02 &  8.478   & 2.43  &  0.17 &    6    \\
V465 Mon     &   0.373  &  6088 &   2.30  &  3.70  & -0.04 &  10.088  & 2.45  &  0.18 &    6    \\
V473 Lyr     &   0.559  &  6113 &   2.60  &  4.50  & -0.06 &  7.726   & 2.35  &  0.15 &    3    \\
V473 Lyr     &   0.793  &  6163 &   2.45  &  4.20  & -0.06 &  7.726   & 2.45  &  0.16 &    3    \\
V482 Sco     &   0.113  &  6129 &   2.20  &  4.00  &  0.07 &  6.945   & 2.31  &  0.13 &    3    \\
V495 Mon     &   0.622  &  5591 &   1.60  &  4.00  & -0.26 &  12.105  & 2.35  &  0.18 &    3    \\
V496 Aql     &   0.043  &  5841 &   1.70  &  4.00  &  0.06 &  6.884   & 2.35  &  0.15 &    6    \\
V496 Aql     &   0.910  &  5822 &   1.70  &  4.25  &  0.03 &  6.884   & 2.35  &  0.16 &    6    \\
V496 Cen     &   0.025  &  6194 &   2.15  &  3.80  &  0.00 &  7.059   & 2.32  &  0.15 &    3    \\
V500 Sco     &   0.051  &  5999 &   1.80  &  4.30  &  0.06 &  6.530   & 2.36  &  0.15 &    6    \\
V500 Sco     &   0.340  &  5356 &   1.25  &  3.60  &  0.02 &  6.530   & 2.33  &  0.16 &    6    \\
V504 Mon     &   0.714  &  6370 &   2.00  &  3.50  & -0.31 &  10.676  & 2.13  &  0.13 &    3    \\
V508 Mon     &   0.742  &  5566 &   1.60  &  4.00  & -0.25 &  10.710  & 2.05  &  0.15 &    3    \\
V510 Mon     &   0.232  &  5842 &   1.40  &  3.80  & -0.19 &  12.956  & 2.48  &  0.15 &    3    \\
\hline                         
\end{tabular}                  
\end{table*}                    

\begin{table*}
\contcaption{}
\begin{tabular}{lccccccccc}
\hline
\hline
Star & phase & T$_{\rm eff}$,\,K & $\log~g$ & V$_{\rm t}$, km~s$^{-1}$ &[Fe/H] &R$_{\rm G}$, (kpc) & 
$\rm \log \epsilon (Ba)_{NLTE}$ & $\sigma (Ba) $ & N lines  \\
\hline
V526 Mon     &   0.674  &  6464 &   2.40  &  3.50  &       &  9.325   & 2.52  &  0.20 &    3    \\
V532 Cyg     &   0.599  &  5960 &   2.00  &  3.10  &  0.08 &  7.981   & 2.47  &  0.15 &    6    \\
V600 Aql     &   0.457  &  5604 &   1.70  &  4.00  & -0.02 &  6.831   & 2.30  &  0.15 &    6    \\
V636 Cas     &   0.151  &  5562 &   1.50  &  4.10  &  0.07 &  8.337   & 2.33  &  0.15 &    6    \\
V636 Cas     &   0.270  &  5473 &   1.50  &  3.80  &  0.06 &  8.337   & 2.39  &  0.15 &    6    \\
V636 Sco     &   0.969  &  5348 &   1.60  &  4.30  &  0.07 &  7.148   & 2.01  &  0.15 &    3    \\
V659 Cen     &   0.856  &  6067 &   2.20  &  5.40  &  0.07 &  7.447   & 2.17  &  0.15 &    3    \\
V733 Aql     &   0.699  &  5435 &   1.70  &  4.70  & -0.01 &  6.187   & 2.02  &  0.15 &    5    \\
V737 Cen     &   0.967  &  5865 &   2.00  &  4.50  &  0.13 &  7.335   & 2.45  &  0.13 &    3    \\
V924 Cyg     &   0.702  &  5910 &   1.80  &  5.00  & -0.09 &  7.529   & 2.10  &  0.20 &    3    \\
V950 Sco     &   0.126  &  6322 &   2.20  &  4.20  &  0.11 &  7.099   & 2.22  &  0.13 &    3    \\
V1154 Cyg    &   0.120  &  5840 &   1.80  &  4.10  & -0.12 &  7.704   & 2.37  &  0.15 &    6    \\
V1334 Cyg    &   0.253  &  6203 &   2.10  &  3.50  &  0.03 &  7.856   & 2.30  &  0.15 &    6    \\
V1334 Cyg    &   0.295  &  6203 &   2.10  &  3.40  &  0.06 &  7.856   & 2.30  &  0.18 &    6    \\
V1726 Cyg    &   0.109  &  6349 &   2.20  &  5.20  & -0.02 &  8.177   & 2.14  &  0.20 &    3    \\
VW Cru       &   0.175  &  5880 &   2.10  &  4.40  &  0.10 &  7.275   & 2.23  &  0.13 &    3    \\
VW Pup       &   0.617  &  5586 &   1.90  &  4.60  & -0.19 &  10.340  & 2.05  &  0.15 &    3    \\
VX Cyg       &   0.411  &  5015 &   0.80  &  3.70  &  0.10 &  8.064   & 2.49  &  0.18 &    6    \\
VX Per       &   0.709  &  5999 &   1.80  &  4.20  & -0.03 &  9.627   & 2.52  &  0.20 &    6    \\
VX Per       &   0.800  &  5992 &   1.80  &  4.20  & -0.04 &  9.627   & 2.52  &  0.18 &    6    \\
VX Pup       &   0.813  &  6159 &   2.50  &  3.50  & -0.13 &  8.638   & 2.30  &  0.20 &    2    \\
VY Car       &   0.169  &  4895 &   1.60  &  5.20  &  0.26 &  7.582   & 2.64  &  0.20 &    3    \\
VY Cyg       &   0.370  &  5743 &   1.80  &  4.10  & -0.02 &  7.885   & 2.27  &  0.15 &    6    \\
VY Sgr       &   0.394  &  5144 &   1.25  &  3.30  &  0.26 &  5.584   & 2.50  &  0.15 &    3    \\
VZ Cma       &   0.777  &  6542 &   2.50  &  3.60  & -0.06 &  8.753   & 2.57  &  0.20 &    2    \\
VZ Cyg       &   0.365  &  5670 &   1.80  &  3.30  & -0.07 &  8.132   & 2.52  &  0.18 &    6    \\
VZ Pup       &   0.980  &  6070 &   2.20  &  4.70  & -0.16 &  10.404  & 2.25  &  0.19 &    3    \\
W Gem        &   0.029  &  6416 &   2.25  &  5.00  & -0.02 &  8.776   & 2.30  &  0.13 &    6    \\
W Gem        &   0.136  &  6032 &   1.85  &  4.05  & -0.01 &  8.776   & 2.35  &  0.14 &    6    \\
W Gem        &   0.524  &  5475 &   1.70  &  3.90  & -0.01 &  8.776   & 2.39  &  0.13 &    6    \\
W Sgr        &   0.302  &  5927 &   1.85  &  3.70  &  0.04 &  7.512   & 2.40  &  0.15 &    6    \\
W Sgr        &   0.462  &  5535 &   1.70  &  3.80  &  0.03 &  7.512   & 2.40  &  0.14 &    6    \\
WW Car       &   0.852  &  5858 &   2.10  &  5.30  & -0.07 &  7.515   & 1.87  &  0.13 &    3    \\
WW Mon       &   0.750  &  6166 &   2.20  &  5.10  & -0.29 &  12.943  & 2.30  &  0.15 &    3    \\
WW Pup       &   0.722  &  5550 &   1.60  &  4.20  & -0.18 &  10.068  & 2.07  &  0.15 &    3    \\
WZ Car       &   0.009  &  5770 &   1.80  &  6.80  &  0.03 &  7.573   & 2.17  &  0.15 &    3    \\
WZ Sgr       &   0.335  &  5077 &   1.10  &  3.80  &  0.20 &  5.964   & 2.40  &  0.15 &    3    \\
WZ Sgr       &   0.942  &  6130 &   1.85  &  5.70  &  0.10 &  5.964   & 2.40  &  0.15 &    3    \\
X Cru        &   0.125  &  5948 &   2.00  &  4.20  &  0.14 &  7.201   & 2.32  &  0.13 &    3    \\
X Lac        &   0.328  &  5890 &   1.85  &  3.30  & -0.01 &  8.477   & 2.42  &  0.18 &    5    \\
X Pup        &   0.175  &  5848 &   1.30  &  5.00  & -0.01 &  9.730   & 2.25  &  0.15 &    6    \\
X Pup        &   0.213  &  5654 &   1.10  &  4.40  &  0.01 &  9.730   & 2.32  &  0.16 &    6    \\
X Vul        &   0.062  &  6260 &   2.00  &  4.20  &  0.13 &  7.532   & 2.17  &  0.15 &    6    \\
X Vul        &   0.220  &  5870 &   1.90  &  4.25  &  0.12 &  7.532   & 2.12  &  0.15 &    6    \\
XX Car       &   0.692  &  5942 &   1.60  &  4.50  &  0.11 &  7.382   & 2.29  &  0.13 &    3    \\
XX Cen       &   0.116  &  5953 &   2.00  &  5.00  &  0.16 &  6.997   & 2.40  &  0.13 &    3    \\
XX Mon       &   0.704  &  5533 &   1.70  &  4.20  & -0.18 &  11.946  & 2.20  &  0.15 &    3    \\
XX Sgr       &   0.584  &  5557 &   1.60  &  3.20  &  0.05 &  6.632   & 2.53  &  0.18 &    6    \\
XX Vel       &   0.011  &  6521 &   2.20  &  4.90  & -0.05 &  7.708   & 2.10  &  0.13 &    3    \\
XY Car       &   0.242  &  5738 &   1.70  &  4.60  &  0.04 &  7.356   & 2.07  &  0.13 &    3    \\
XY Cas       &   0.603  &  5542 &   1.50  &  3.80  &  0.02 &  8.979   & 2.28  &  0.15 &    5    \\
XZ Car       &   0.072  &  6170 &   2.10  &  5.70  &  0.14 &  7.414   & 2.32  &  0.15 &    1    \\
Y Lac        &   0.858  &  6006 &   1.70  &  4.45  & -0.08 &  8.419   & 2.17  &  0.18 &    6    \\
Y Lac        &   0.936  &  6330 &   2.00  &  4.00  & -0.10 &  8.419   & 2.17  &  0.17 &    6    \\
Y Sgr        &   0.319  &  5733 &   1.65  &  3.80  &  0.11 &  7.425   & 2.43  &  0.18 &    6    \\
Y Sgr        &   0.427  &  5605 &   1.60  &  3.70  &  0.04 &  7.425   & 2.45  &  0.18 &    6    \\
YZ Car       &   0.860  &  5655 &   1.60  &  6.50  &  0.02 &  7.629   & 2.12  &  0.13 &    3    \\
YZ Sgr       &   0.249  &  5507 &   1.40  &  3.65  &  0.06 &  6.799   & 2.24  &  0.15 &    6    \\
YZ Sgr       &   0.775  &  5913 &   1.90  &  4.80  &  0.08 &  6.799   & 2.27  &  0.15 &    6    \\
Z Lac        &   0.204  &  5722 &   1.50  &  3.80  &  0.05 &  8.564   & 2.37  &  0.15 &    6    \\
Z Lac        &   0.930  &  5899 &   1.70  &  4.30  &  0.01 &  8.564   & 2.37  &  0.15 &    6    \\
\hline   
\end{tabular}
\end{table*}

\label{lastpage}

\bsp


\begin{thebibliography}{}


\bibitem[\protect\citeauthoryear{Andrievsky et al.}{2002a}]{An02a}
   Andrievsky S.M.,  Kovtyukh V.V., Luck R.E., L\'epine J.R.D., Bersier D., 
   Maciel W.J., Barbuy B., Klochkova V.G., Panchuk V.E., Karpischek R.U., 2002a,
   A\&A 381, 32
  
\bibitem[\protect\citeauthoryear{Andrievsky et al.}{2002b}]{An02b}
   Andrievsky S.M., Bersier D., Kovtyukh V.V., Luck R.E., Maciel W.J., 
   L\'epine J.R.D., Beletsky Yu.V., 2002b, A\&A 384, 140

\bibitem[\protect\citeauthoryear{Andrievsky et al.}{2002c}]{An02c}
   Andrievsky S.M., Kovtyukh V.V., Luck R.E., L\'epine J.R.D., Maciel W.J., 
   Beletsky Yu.V., 2002c, A\&A 392, 491

\bibitem[\protect\citeauthoryear{Andrievsky et al.}{2005}]{An05}
   Andrievsky S.M., Luck R.E., Kovtyukh V.V.,
   2005, AJ 130, 1880.

\bibitem[\protect\citeauthoryear{Andrievsky et al.}{2004}]{An04}
   Andrievsky S.M., Luck R.E., Martin P., L\'epine J.R.D., 2004, A\&A 413, 159

\bibitem[\protect\citeauthoryear{Andrievsky et al.}{2009}]{An09}
   Andrievsky S.M., Spite M., Korotin S.A., Spite F., Fran\,cois P., Bonifacio P.,
   Cayrel R., Hill V., 2009, A\&A 494, 1083

\bibitem[\protect\citeauthoryear{Cameron}{1982}]{Cam82}
   Cameron A.G.W., 1982, ApSS 82, 123 

\bibitem[\protect\citeauthoryear{Carlsson}{1986}]{Ca86}
   Carlsson M., 1986, Uppsala Obs. Rep. 33   

\bibitem[\protect\citeauthoryear{Cescutti et al.}{2007}]{Ce07}
   Cescutti G., Matteucci F., Fran\,cois P., Chiappini C., 2007, A\&A 462, 943


\bibitem[\protect\citeauthoryear{Fuhr \& Wiese}{2006}]{FurWi06}
    Fuhr J.R. \& Wiese W.L., 2006, J. Phys. Chem. Ref. Data 35, 1669.

\bibitem[\protect\citeauthoryear{Korotin et al.}{1999}]{Ko99}
   Korotin S.A., Andrievsky S.M., Luck R.E., 1999, A\&A 351, 168

\bibitem[\protect\citeauthoryear{Korotin et al.}{2010}]{Ko10}
   Korotin S., Mishenina T., Gorbaneva T., Soubiran C., 2010, Proceedings of the 
   11th Symposium on Nuclei in the Cosmos. 19-23 July 2010. Heidelberg, Germany

\bibitem[\protect\citeauthoryear{Kovtyukh et al.}{2005}]{Ko05}
   Kovtyukh, V.V., Andrievsky S.M., Belik S.I., Luck R.E.,
   2005, AJ 129, 433

 
\bibitem[\protect\citeauthoryear{L\'epine et al.}{2011}]{Le11}
     L\'epine J.R.D., Cruz P., Scarano S., Jr., Barros D.A., Dias W.S., 
     Pompeia L., Andrievsky S.M., Carraro G., Famaey B., 2011, MNRAS 417, 698

\bibitem[\protect\citeauthoryear{Luck \& Andrievsky}{2004}]{Lu04}
   Luck R.E., Andrievsky S.M.,
   2004, AJ 128, 343

\bibitem[\protect\citeauthoryear{Luck et al.}{2008}]{Lu08}
   Luck, R.E., Andrievsky S.M., Fokin, A. \& Kovtyukh V.V.,
   2008, AJ 136, 98.

\bibitem[\protect\citeauthoryear{Luck et al.}{2011}]{Lu11}
   Luck R.E., Andrievsky S.M., Kovtyukh V.V., Gieren W., Graczyk D.,
   2011, AJ 142, 51L

\bibitem[\protect\citeauthoryear{Luck et al.}{2003}]{Lu03}
   Luck R.E., Gieren W.P., Andrievsky S.M., Kovtyukh V.V., Fouqu\'e P., 
   Pont F., Kienzle F., 2003, A\&A 401, 939

\bibitem[\protect\citeauthoryear{Luck et al.}{2006}]{Lu06}
   Luck R.E., Kovtyukh V.V., Andrievsky S.M., 2006, AJ 132, 902
   
\bibitem[\protect\citeauthoryear{Luck \& Lambert}{2011}]{LL11}
   Luck R.E., Lambert, D.L.,
   2011, AJ 142, 137L


\bibitem[\protect\citeauthoryear{Mashonkina et al.}{1999}]{Mashet99}
    Mashonkina L., Gehren T., Bikmaev I., 1999, A\&A 343, 519

\bibitem[\protect\citeauthoryear{Rutten}{1978}]{Rut78}
    Rutten R.J., 1978, Solar Phys. 56, 237

\bibitem[\protect\citeauthoryear{Simmerer et al.}{2004}]{Si04}
   Simmerer J., Sneden Ch., Cowan J.J. et al., 2004, ApJ 617, 1091 

\bibitem[\protect\citeauthoryear{Stodilka \& Malynch}{2006}]{St06}
   Stodilka, M.I., Malynch, S.Z,
   2006, MNRAS 373, 1523


\bibitem[\protect\citeauthoryear{Tsymbal}{1996}]{Ts96}
   Tsymbal V.V., 1996, Model Atmospheres and Spectrum Synthesis, ed. S.J. Adelman, F. Kupka,  
   W.W. Weiss (San Francisco), ASP Conf. Ser., 108

\end{thebibliography}
\end{document}